\documentclass[
 reprint,
 superscriptaddress,
showpacs,preprintnumbers,
 amsmath,amssymb,
 aps,
pra
]{revtex4-1}

\usepackage{amsmath,amssymb,bbm,epsfig}

\usepackage{color}
\usepackage[usenames,dvipsnames,svgnames,table]{xcolor}
\usepackage[colorlinks=true,
            linkcolor=blue,
            urlcolor=blue,
            citecolor=blue]{hyperref}

\usepackage[abs]{overpic}
\usepackage{enumerate}

\newcommand{\be}{\begin{equation}}
\newcommand{\ee}{\end{equation}}
\newcommand{\ba}{\begin{array}}
\newcommand{\ea}{\end{array}}
\newcommand{\bqa}{\begin{eqnarray}}
\newcommand{\eqa}{\end{eqnarray}}

\newcommand{\ket}[1]{\ensuremath{| #1 \rangle}}

\newcommand{\id}{{\mathbbm 1}}

\begin{document}

\title{Quasi-periodically driven quantum systems}

\author{Albert Verdeny}
\affiliation{Department of Physics, Imperial College London, London SW7 2AZ, United Kingdom}
\author{Joaquim Puig}
\affiliation{Department of Mathematics, Universitat Polit\`ecnica de Catalunya, 08007 Barcelona, Spain}
\author{Florian Mintert}
\affiliation{Department of Physics, Imperial College London, London SW7 2AZ, United Kingdom}

\begin{abstract}

Floquet theory provides rigorous foundations for the theory of periodically driven quantum systems.
In the case of non-periodic driving, however, the situation is not so well understood.
Here, we provide a critical review of the theoretical framework developed for quasi-periodically driven quantum systems.
Although the theoretical footing is still under development,
we argue that
quasi-periodically driven quantum systems can be treated with generalizations of Floquet theory
in suitable parameter regimes. Moreover, we provide a generalization of the Floquet-Magnus expansion and argue that quasi-periodic driving offers a promising route for quantum simulations.

\end{abstract}

\maketitle

\section{Introduction}

The dynamics of quantum systems induced by a time-dependent Hamiltonian attracts attention from various communities.
Chemical reactions can be controlled with driving induced by laser beams \cite{shapiro:187},
and driving atoms permits to investigate their electronic structure \cite{RevModPhys.84.1011}.
Suitably chosen driving sequences permit to investigate dynamics in macro-molecular complexes \cite{engel} and there exist phases in solid state systems that can be accessed only in the presence of driving \cite{Ichikawa:2011zf,doi:10.1143/JPSJ.75.011001}.
A neat bridge between quantum optics and solid state physics is built by the fact that periodically driven atomic gases can be employed as quantum simulators for models of solid state theory \cite{Bukov14ar,Goldman14}.

Solving the Schr\"odinger equation with a time-dependent Hamiltonian calls 
for different mathematical techniques than those applied in situations with time-independent Hamiltonians.
Differential equations with time-dependent coefficients have been investigated thoroughly
and, in particular, developments regarding reducibility are appreciated, since they permit to understand driven systems in terms of time-independent Hamiltonians \cite{eliasson:floquet}.

The foundation for this is laid by the Floquet theorem \cite{Floquet83}, which 
relates a periodically time-dependent with a constant Hamiltonian.
This mathematical theorem provides the basis for experiments that employ periodically driven quantum systems for quantum simulations of systems with time-independent Hamiltonians.
Such experiments have led to the experimental observation of e.g. coherent suppression of tunneling \cite{Keay95,Madison98,Lignier07}, spin-orbit coupling \cite{Lin11,Jotzu15}, synthetic magnetism \cite{Struck12,Aidelsburger13,Miyake13}, ferromagnetic domains \cite{Parker13}, or topological band structures \cite{Jotzu14,Aidelsburger15}

The specific time-dependence of the driving force plays a crucial role in the dynamics that driven systems can undergo. Yet, despite the possibility to experimentally tune it, very simple driving protocols are usually employed, which can significantly limit the performance of the simulations \cite{Verdeny14} and restrict the range of accessible dynamics \cite{Hauke12,Verdeny15}.

In this context, pulse-shaping techniques have been introduced in order to achieve the simulation of the desired effective dynamics in an optimal fashion \cite{Verdeny14,Verdeny15}.
Yet, the restriction to periodic driving is a limitation,
and quasi-periodic driving,  i.e. driving with a time-dependence characterized by several frequencies that can be irrationally related, promises substantially enhanced control over the quantum system at hand.
Since the use of quasi-periodic driving \cite{EPJD.26.219,PhysRevLett.96.240604,PhysRevE.86.056201}, however, implies that Floquet theorem is not applicable,
the mathematical foundation is far less solid than in the case of periodic driving.

Generalizations of Floquet's theorem to quasi-periodic driving have been pursued both in the quantum physics/chemistry literature \cite{MMFT,Chu04} and in the mathematical literature \cite{Jorba1992a,eliasson:floquet,Krikorian2014,Avila2006}.
The former perspective approaches quasi-periodic driving as a limiting case of periodic systems, while the mathematical literature approaches quasi-periodicity without resorting to results from periodic systems.
Beyond the fundamentally different approaches, also the findings in the different communities are not always consistent with each other.

The goal of the present article is twofold. On the one hand, we discuss prior literature on the generalization of Floquet's theorem to quasi-periodic systems, and attempt to overview over what findings have been verified to mathematical rigour, and what findings are rather based on case studies and still lack a general, rigorous foundation. On the other hand, we aim at studying the possibility to use quasi-periodically driven systems for quantum simulations.

We consider quasi-periodic Hamiltonians $H(t)$ that can be defined in terms of a Fourier-like representation of the form
\begin{eqnarray}\label{qpham}
H(t)=\sum_\textbf{n} H_{\textbf{n}} e^{i \textbf{n}\cdot \boldsymbol{\omega} t},
\end{eqnarray}
where $\boldsymbol{\omega}=(\omega_1,\cdots,\omega_d)$ 
is a finite-dimensional vector of real frequencies that are irrationally related, and  $\textbf{n}=(n_1,\cdots, n_d)$
is a vector of integers such that $\textbf{n}\cdot \boldsymbol{\omega}= n_1 \omega_1+\cdots+n_d\omega_d$. Moreover, the norm of coefficients $H_\textbf{n}$ are considered to decay sufficiently fast with $|\textbf{n}|$.

The main underlying question of the present work is the possibility to express the time-evolution operator $U(t)$ of a quasi-periodically driven system in terms of a generalized Floquet representation of the form
\begin{eqnarray}
U(t)\stackrel{?}{=}U_Q^\dagger(t)e^{-i H_Q t}, \label{decompqp}
\end{eqnarray}
where $H_Q$ is a time-independent Hermitian operator and $U_Q(t)=\sum_\textbf{n} U_{\textbf{n}} e^{i \textbf{n}\cdot \boldsymbol{\omega} t}$ is a quasi-periodic unitary characterized by the same fundamental frequency vector $\boldsymbol{\omega}$ as the quasi-periodic Hamiltonian $H(t)$. 
If the frequency vector $\boldsymbol{\omega}$ defining the quasi-periodicity of $H(t)$ contains only one element, i.e. $d=1$, the Hamiltonian becomes periodic with period $2\pi/\omega_1$ and the decomposition in Eq. \eqref{decompqp} reduces to the usual Floquet representation, which is known to exist. 
However, if $\boldsymbol{\omega}$ contains more than one element, i.e. $d>1$, the Hamiltonian is not periodic
and the representation in Eq. \eqref{decompqp} is \textit{a priori} not guaranteed. 

The possibility to represent the time-evolution operator of a quasi-periodically driven system as in Eq. \eqref{decompqp} is directly related to the problem of reducibility \cite{Eliasson1998,eliasson2006linear} of first-order differential equations with quasi-periodic coefficients, which is still an ongoing problem in the mathematics community \cite{karaliolios:global}. Unlike their periodic counterparts, linear differential equations with quasi-periodic  coefficients  cannot  always  be  reduced  to  constant  coefficients  by means of a quasi-periodic transformation \cite{eliasson:so3,krikorian:c0}, although a quasi-periodic Floquet reducibility theory does exist \cite{Jorba1992a,her-you}.

Generalizations of Floquet theory to quasi-periodically driven systems have been derived also from a less mathematical perspective.
Many-mode Floquet theory (MMFT) \cite{MMFT,MMFTii,MMFTiii,MMFTiv} is based on physical assumptions of the underlying time-dependent Hamiltonian, and it has been successfully applied to a variety of systems ranging from quantum chemistry \cite{MMFT,Leskes07} to quantum optics \cite{Son08,Zhao15}. However, it does not seem to have an entirely rigorous footing yet.

In this article, we address these different perspectives and argue that, despite gaps in a general mathematical footing, concepts from regular Floquet theory can be translated directly to quasi-periodically driven systems, especially in fast-driving regimes,  i.e. the regime of quantum simulations.

In Sec. \ref{sec:Floq} we introduce notation and preliminary concepts of Floquet theory that will be used throughout the article. 
In Sec. \ref{sec:MMFTR}, we revise critically the MMFT and point out aspects of the derivation that cast doubts on the general validity of the proof. 
In Sec. \ref{sec:QPFourier} we argue how, nevertheless, the general formalism of MMFT can still lead to valid results, in agreement with prior work \cite{Chu04,MMFTii,MMFTiii,MMFTiv,Son08,Forney10,Zhao15}.
In Sec. \ref{sec:High}, we derive a generalization of the Floquet-Magnus expansion \cite{Casas01}, which provides a perturbative exponential expansion of the time-evolution operator that has the desired structure.
With this, we advocate the possibility to identify effective Hamiltonians that characterize well the effective dynamics of quasi-periodically driven systems in a fast-driving regime, and exemplify in Sec. \ref{sec:Lambda} the results with a quasi-periodically driven Lambda system.

\section{Floquet theory} \label{sec:Floq}

Floquet's theorem \cite{Floquet83} asserts that the Schr\"odinger equation
\begin{eqnarray}\label{Schrod}
i\partial_t \tilde{U}(t)=\tilde{H}(t)\tilde{U}(t),
\end{eqnarray} 
characterizing the time-evolution operator $\tilde{U}(t)$ of a system described by a periodic Hamiltonian $\tilde{H}(t)=\tilde{H}(t+T)$, is \textit{reducible}. That is, there exists a periodic unitary $U_P(t)=U_P(t+T)$ that transforms the Schr\"odinger operator $\tilde{K}(t)=\tilde{H}(t)-i\partial_t$ into
\begin{eqnarray}\label{Ktrans}
U_P(t)\tilde{K}(t)U_P^\dagger(t)=H_F-i\partial_t\ ,
\end{eqnarray}
where $H_F=U_P(t)\tilde{H}(t)U_P^\dagger(t)-iU_P(t)(\partial_tU_P^\dagger(t))$ is a time-independent Hamiltonian \footnote{The periodic unitary $U_P(t)$ and the Hamiltonian $H_F$ are, however, not uniquely defined. In particular, the eigenvalues of $H_F$ are only defined up to multiples of $\omega$.}. 
As a consequence, the time-evolution operator of the system can be represented as the product 
\begin{eqnarray}
\tilde{U}(t)=U_P^\dagger(t)e^{-iH_F t}, \label{decomp}
\end{eqnarray}
with $U_P(0)=\id$. This decomposition is of central importance in the context of quantum simulations with periodically driven systems because it ensures that, in a suitable fast-driving regime, the dynamics of the driven system can be approximated in terms of the time-independent Hamiltonian $H_F$
\cite{Bukov14ar,Goldman14}.

The eigenstates $|\epsilon_k\rangle$ of the Hamiltonian $H_F$ form a basis in the system Hilbert space $\mathcal{H}$, on which the periodic Hamiltonian $\tilde{H}(t)$ acts. Thus, any vector $|\phi(0)\rangle$ characterizing the initial state of the system can be written as a linear combination of the eigenstates $|\epsilon_k\rangle$.
Consequently, the decomposition of the time-evolution operator in Eq. (\ref{decomp}) implies that time-dependent states $|\phi(t)\rangle=\tilde{U}(t)|\phi(0)\rangle$ can be expressed as a linear combination with time-independent coefficients of Floquet states of the form
\begin{eqnarray}\label{floquetstates}
|\phi_k(t)\rangle=e^{-i\epsilon_k t} |u_k(t)\rangle,
\end{eqnarray}
where $\epsilon_k$ are the eigenvalues of $H_F$ (also termed quasienergies), and $|u_k(t)\rangle=U^\dagger_P(t)|\epsilon_k\rangle=|u_k(t+T)\rangle$ are periodic state vectors.

The quasienergies $\epsilon_k$ play a very important role in the dynamics of driven systems.
They can be calculated after inserting the Floquet states in Eq. \eqref{floquetstates} into the 
Schr\"odinger equation $i\partial_t |\phi_k(t)\rangle=\tilde{H}(t)|\phi_k(t)\rangle$, which yields 
\begin{eqnarray}\label{eigenvalue}
\tilde{K}(t) |u_k(t)\rangle=\epsilon_k |u_k(t)\rangle\ .
\end{eqnarray}
Eq. \eqref{eigenvalue} formally describes an eigenvalue problem resembling the time-independent Schr\"odinger equation, where the periodic states $|u_k(t)\rangle$ play an analogous role of stationary states.
Due to the time-dependence of the states and the action of the derivative in $\tilde{K}(t)$, however, the diagonalization in Eq. \eqref{eigenvalue} cannot be straightforwardly solved with standard matrix diagonalization techniques.

For this reason, it is often convenient to formulate the problem in a Fourier space where the operator $\tilde{K}(t)$ is treated as an infinite-dimensional time-independent operator \cite{Shirley65,Sambe73}.

\subsection{Time-independent formalism}\label{sec:timeindep}

State vectors of the driven system are defined on the system Hilbert space $\mathcal{H}$, where time is regarded as a parameter. 
In order to arrive at a formalism in which the parameter `time' does not appear explicitly,
one exploits the fact that the states $|u(t)\rangle$ in $\mathcal{H}$ that have a periodic time-dependence
can be defined on a Floquet Hilbert space $\mathcal{F}=\mathcal{H}\otimes L^2(\mathbb{T})$, where $L^2(\mathbb{T})$ is the Hilbert space of periodic functions \cite{Sambe73}.
In this Floquet space, time is not regarded as a parameter but rather as a coordinate of the new Hilbert space. 
The explicit time-dependence of the system can then be removed by adopting a Fourier representation of the periodic states in the space $\mathcal{F}$.
Fourier series permit the characterization of periodic functions in terms of a sequence of its Fourier coefficients. 
Formally, this can be described through an isomorphism between the space $L^2(\mathbb{T})$ of periodic functions and the space $l^2(\mathbb{Z})$ of square-summable sequences. 
This isomorphism allows one to map the exponential functions $e^{in\omega t}$, which form a basis in $L^2(\mathbb{T})$, to states $|n\rangle$, which define an orthonormal basis in $l^2(\mathbb{Z})$. 

In this manner, periodic states $|u(t)\rangle=\sum_n|u_n\rangle e^{in\omega t}$  are mapped to states
\begin{eqnarray}
|u\rangle=\sum_n|u_n\rangle \otimes |n\rangle,
\end{eqnarray}
while periodic operators $A(t)=\sum_n A_ne^{in\omega t}$ can be mapped to
\begin{eqnarray}\label{PerOper}
\mathcal{A}=\sum_n A_n\otimes \sigma_n,
\end{eqnarray}
were the ladder operators $\sigma_n =\sum_{m}|m+n\rangle\langle m|$ satisfy $\sigma_n|m\rangle=|n+m\rangle$.
Similarly, the derivative operator $-i\partial_t$ is mapped to
\begin{eqnarray}
\label{Doperator}
\tilde{\mathcal{D}}=\id\otimes\omega \hat{n},
\end{eqnarray}
with the number operator $\hat{n}=\sum_n n|n\rangle\langle n|$ satisfying $\hat{n}|n\rangle=n|n\rangle$ and the commutation relation $[\hat{n},\sigma_m]=m\sigma_m$.
 
Consequently, the isomorphism between $\mathcal{F}$ and $\mathcal{H}\otimes l^2(\mathbb{Z})$ 
permits one to treat the periodic system within a Fourier formalism by mapping the Schr\"odinger operator $\tilde{K}(t)=\tilde{H}(t)-i\partial_t$ to the operator \cite{Verdeny13}
\begin{eqnarray}\label{K}
\tilde{\mathcal{K}}=\sum_n H_n\otimes \sigma_n +\id\otimes \omega \hat{n}\ .
\end{eqnarray}
The time-evolution operator $\tilde{U}(t)$ of the system can then be calculated via the relation \cite{Shirley65}
\begin{eqnarray}\label{UFloq}
\tilde{U}(t)=\sum_n \langle n |e^{-i\tilde{\mathcal{K}}t}|0\rangle e^{in\omega t}\ ,
\end{eqnarray}
which can be readily verified by inserting Eq. \eqref{UFloq} into the Schr\"odinger equation and using the explicit form of $\tilde{\mathcal{K}}$ given in Eq. \eqref{K}, as we explicitely demonstrate in Appendix \ref{appendix} for illustrative purposes.

The operator $\tilde{\mathcal{K}}$ in Eq. \eqref{K} is often represented as an infinite-dimensional matrix.
The Floquet states in Eq. (\ref{floquetstates}) can be obtained from the diagonalization of $\tilde{\mathcal{K}}$ \cite{Shirley65}. 
In order to find the Floquet decomposition of the time-evolution operator in Eq. (\ref{decomp}), however, it is not necessary to completely diagonalize the operator $\tilde{\mathcal{K}}$. Instead, the operator $\tilde{\mathcal{K}}$ needs to be brought into the block-diagonal form
\begin{eqnarray}\label{blockdiagFloq}
\tilde{\mathcal{K}}_{B}= \mathcal{U}_P\tilde{\mathcal{K}}\mathcal{U}^\dagger_P=H_F\otimes \id +\id\otimes \omega \hat{n}
\end{eqnarray}
by means of a unitary transformation
\begin{eqnarray}\label{UpFloq}
\mathcal{U}_P=\sum_n U_n\otimes \sigma_n.
\end{eqnarray}
The block-diagonalization in Eq. \eqref{blockdiagFloq} describes the counterpart in the present time-independent formalism of the transformation in Eq. (\ref{Ktrans}) such that, if the block-diagonalization is achieved, the Floquet Hamiltonian $H_F$ and the periodic unitary ${U}_P(t)=\sum_n U_n e^{in\omega t}$ are straightforwardly obtained from Eqs.  \eqref{blockdiagFloq} and \eqref{UpFloq}. 

\section{Many-mode Floquet theory revised} \label{sec:MMFTR}

The many-mode Floquet theory (MMFT) \cite{MMFT,MMFTii,MMFTiii,MMFTiv} was introduced in the context of quantum chemistry  as a generalization of Floquet theory to treat systems with a quasi-periodic time dependence. 
The derivation of MMFT is rooted on Floquet's theorem and its proposed generality contrasts with other results derived with more rigorous approaches. 
In this section, we revise the derivation of MMFT and challenge aspects of the proof that question its general validity.

The derivation of the MMFT \cite{MMFT} consists in approximating the quasi-periodic Hamiltonian $H(t)$ by a periodic Hamiltonian and, then, using Floquet theory to demonstrate the existence of a generalized Floquet decomposition for the time-evolution operator of the system. The derivation \cite{MMFT} starts by considering a quasi-periodic Hamiltonian $H(t)=\sum_\textbf{n} H_{\textbf{n}} e^{i \textbf{n}\cdot \boldsymbol{\omega} t}$ in Eq. \eqref{qpham} and approximating the different elements $\omega_i$ of the frequency vector $\boldsymbol{\omega}$ by a fraction.
Then, a small fundamental driving frequency $\omega$ is identified such that the different irrationally-related frequencies $\omega_i$ are expressed as
\begin{eqnarray}\label{approx}
\omega_i\approx N_i\ \omega,
\end{eqnarray}
with some integers $N_i$.
In this manner,  the quasi-periodically driven Hamiltonian $H(t)$ can be approximated by the periodic Hamiltonian
\begin{eqnarray}\label{approxH}
\tilde{H}(t)=\sum_\textbf{n} H_\textbf{n} e^{i \textbf{n}\cdot \textbf{N}\omega t},
\end{eqnarray}
where $\textbf{N}=(N_1,\cdots,N_d)$. The validity of the approximation $H(t)\approx \tilde{H}(t)$ for a certain time window importantly depends on the good behavior of the Hamiltonian and on the approximation in Eq. \eqref{approx}, which can be performed with any desired accuracy.
When the approximation $H(t)\approx \tilde{H}(t)$ is satisfied with sufficient accuracy, the time-evolution operator $U(t)$ of the quasi-periodically driven system can be also approximated by the time-evolution operator $\tilde{U}(t)$ that is induced by the periodic approximated Hamiltonian $\tilde{H}(t)$, i.e. $U(t)\approx\tilde{U}(t)$.

The next step in the derivation aims at demonstrating that time-evolution operator $\tilde{U}(t)$ of the periodic Hamiltonian $\tilde{H}(t)$ can be approximately represented by a generalized Floquet decomposition analogous to Eq. \eqref{decompqp}. Specifically, the aim is to express the periodic unitary $U_P(t)$ of the Floquet decomposition in Eq. \eqref{decomp} 
in terms of a Fourier series of the form 
\begin{eqnarray}\label{desiredUp}
U_P(t)=\sum_\textbf{n} U_\textbf{n} e^{i \textbf{n}\cdot \textbf{N}\omega t},
\end{eqnarray} 
which contains only specific Fourier components, since, in general, not all integers can be expressed as $\textbf{n}\cdot \textbf{N}$ with a vector of integers $\textbf{n}$.
If this was possible for an arbitrarily small frequency $\omega$, the unitary $U_P(t)$ in Eq. \eqref{desiredUp} would approximate a quasi-periodic unitary $U_Q(t)=\sum_\textbf{n} U_\textbf{n} e^{i \textbf{n}\cdot \boldsymbol{\omega} t}$ and  the time-evolution operator of the quasi-periodically driven system could 
be well approximated by the sought generalized Floquet decomposition in Eq. \eqref{decompqp}.

The MMFT derivation \cite{MMFT} considers, for concreteness, a quasi-periodic Hamiltonian $H(t)$
in Eq. \eqref{qpham} with $d=2$; that is, the frequency vector contains only two components: $\omega_1$ and $\omega_2$. Moreover, the only non-vanishing coefficients $H_\textbf{n}$ of the quasi-periodic Hamiltonian are $H_{(0,0)}$, $H_{(\pm1,0)}$, and $H_{(0,\pm1)}$. Despite this specific choice, however, the possibility to generalize the results to Hamiltonians containing more frequencies is claimed.

In order to demonstrate the possibility to write the unitary in Eq. \eqref{desiredUp}, the derivation in \cite{MMFT} makes use of the time-independent Floquet formalism described in Sec. \ref{sec:Floq}.
First of all, the operator $\tilde{\mathcal{K}}$ in Eq. \eqref{K} is defined (using a slightly different notation) for the approximated periodic Hamiltonian $\tilde{H}(t)$ in Eq. \eqref{approxH}  
\footnote{For convenience, we use the notation introduced in Sec. \ref{sec:Floq}, which differs from the one used in \cite{MMFT}. In particular, the operator $\tilde{\mathcal{K}}$ is denoted by $H_F$ in \cite{MMFT} and it is represented in terms of a matrix.}.
Then, a block-diagonal  structure is given to $\tilde{\mathcal{K}}$ as a first step to achieve the desired structure of the time-evolution operator. 

The block-diagonal structure is obtained by `relabelling' each vector $|n\rangle$ that forms a basis of $l^2(\mathbb{Z})$ (introduced in Sec. \ref{sec:timeindep}) as
\begin{eqnarray}\label{relabel}
|\textbf{n}p\rangle,
\end{eqnarray}
where the vector of integers $\textbf{n}=(n_1, n_2)$ and the integer $p$ are found by solving the Diophantine equation
\begin{eqnarray}\label{dioph}
n=\sum_{i=1}^2 n_iN_i+p
\end{eqnarray}
for all $n$ and for the integers $N_i$ in Eq. \eqref{approx}.
Thereafter, a tensor product structure  is given to the Hilbert space $l^2(\mathbb{Z})$, such that the state vectors in Eq. \eqref{relabel} are written in the tensor product form $|\textbf{n}p\rangle=|\textbf{n} \rangle |p\rangle$, with $|\textbf{n}\rangle =|n_1\rangle |n_2\rangle $ and where $n_1$, $n_2$, and $p$ can take all integer values. 
In this manner, the operator $\tilde{\mathcal{K}}$ in Eq. \eqref{K} is 
 described to be rewritten in the block-diagonal form
\begin{eqnarray}\label{Ktensor}
\tilde{\mathcal{K}}_{bd}=\mathcal{K} \otimes\id+\id\otimes\omega\hat{p},
\end{eqnarray}
with
\begin{eqnarray}\label{Kq}
\mathcal{K}&=&\sum_\textbf{n} H_\textbf{n}\otimes \sigma_\textbf{n}+\id\otimes \boldsymbol{\omega}\cdot\hat{\textbf{n}}.
\end{eqnarray}
The ladder operator $\sigma_\textbf{n}$ and number operators $\hat{\textbf{n}}$ and $\hat{p}$ introduced in Eq. \eqref{Kq} are defined as $\sigma_\textbf{n}=\sum_{\textbf{m}}|\textbf{m}+\textbf{n}\rangle\langle \textbf{m}|$, $\hat{\textbf{n}}=\sum_\textbf{n} \textbf{n} |\textbf{n}\rangle\langle \textbf{n}|$, and $\hat{p}=\sum_p p |p\rangle\langle p|$, respectively, where the summations include all possible values of $\textbf{n}$ and $p$.

The notation of the states $|n\rangle$ introduced in Eq. \eqref{relabel}, and the tensor structure given to the them and to the operator $\tilde{\mathcal{K}}$ in Eq. \eqref{Ktensor} is of central importance in the derivation of MMFT and is the main focus of our criticism.

A linear Diophantine equation of the form in Eq. \eqref{dioph} with unknowns $p$ and $n_i$ can always be solved independently of the specific integer values $n$ and $N_i$ \footnote{A Diophantine equation $n=\sum_{i=1}^d n_iN_i$ with unknowns $n_i$ can be solved if and only if the greatest common divisor $\mbox{gcd}(N_1,\cdots,N_d)$ divides $n$  \cite{Mordell}. Thus, by appropriately choosing the variable $p$, the Diophantine equation in Eq. \eqref{dioph} can always be solved}. In fact, it has infinitely many solutions. For instance, given a solution $\{\textbf{n}p\}$ it is always possible to obtain another solution by redefining the vector $\textbf{n}$ as $\textbf{n}'=(n_1+zN_2,n_2-zN_1)$ with an arbitrary integer $z$.
For this reason, it is not possible to uniquely associate a single vector $|\textbf{n}p\rangle$ with each vector $|n\rangle$ without a specific prescription of which solution to choose.  Such prescription, however, is not given in \cite{MMFT} and is not compatible with the tensor structure provided \cite{MMFT}.

Problems arising from the ambiguity in the identification of the vector $\ket{{\bf n}p}$ in Eq. \eqref{relabel} become apparent when considering e.g. the scalar product of two states $|\textbf{n}p\rangle$ and $|\textbf{n}'p'\rangle$ that correspond to two solutions $\{\textbf{n}p\}$ and $\{\textbf{n}'p'\}$. 
The scalar product $\langle \textbf{n}p|\textbf{n}'p'\rangle$ vanishes if the two solutions are different. However, with the original notation both states are associated with the same state $|n\rangle$ and the corresponding scalar product $\langle n|n\rangle$ does not vanish, which leads to an inconsistency. 
This problem becomes especially relevant when considering the expression of the operator $\tilde{\mathcal{K}}$ in Eq. \eqref{Ktensor}, which contains infinitely many different matrix elements $\langle \textbf{n}p|\tilde{\mathcal{K}}|\textbf{m}q\rangle$ that correspond to the same matrix element $\langle n|\tilde{\mathcal{K}}|m\rangle$ of the operator $\tilde{\mathcal{K}}$ in Eq. \eqref{K}. 
That is, the operator $\tilde{\mathcal{K}}_{bd}$ in Eq. \eqref{Ktensor} is in fact \textit{not} a mere rewritten version of $\tilde{\mathcal{K}}$ in Eq. \eqref{K} but rather a different operator. 

A central step in the derivation of MMFT is the existence of a unitary transformation relating the operators $\tilde{\mathcal{K}}_{bd}$
defined in Eq. \eqref{Ktensor} and
\be
\tilde{\mathcal{K}}_d=\mathcal{K}_B\otimes\id+\id\otimes\omega\hat{p}
\label{eq:Ktransform}
\ee
with
\be \label{KB}
\mathcal{K}_B=H_F\otimes \id +\id\otimes \boldsymbol{\omega}\cdot \hat{\textbf{n}}
\ee
The existence would follow from a bi-jective relation between $\ket{n}$ and $\ket{{\bf n}p}$;
but since such a relation does not exist, the unitary equivalence between the two operators does not necessarily hold true.

The notation introduced in Eq. \eqref{relabel} is also employed to express the time-evolution operator $\tilde{U}(t)$ in Eq. \eqref{UFloq} as
\begin{eqnarray}\label{Uwrong}
\sum_{n_1,n_2, p=-\infty}^\infty \langle n_1 n_2 p |e^{-i\tilde{\mathcal{K}}t}|0 0 0\rangle e^{in\omega t}.
\end{eqnarray}
This expression, however, contains infinitely many duplicate terms
since there are infinitely many vectors $\langle  n_1 n_2 p |$ that correspond to the same vector $\langle  n |$, according to the prescription given by the Diophantine equation in Eq. \eqref{dioph}. 
Eq.~\eqref{Uwrong} is thus not a reformulation of the expression of the time-evolution operator in Eq. \eqref{UFloq}.

The derivation of MMFT \cite{MMFT} achieves the desired structure of the time-evolution operator by combining the expression for the time-evolution operator $\tilde{U}(t)$ in Eq. \eqref{Uwrong} with the expression for the operator $\tilde{\mathcal{K}}$ in Eq. \eqref{Ktensor}.
Given the doubts on the unitary equivalence between $\tilde{\mathcal{K}}_{bd}$ and $\tilde{\cal{K}}_d$ and the correctness of Eq.~\eqref{Uwrong}, it seems to us that the derivation of MMFT is not complete.

Besides a Floquet-like decomposition for quasi-periodic systems, MMFT also describes a method to calculate the time-evolution operator by diagonalizing a time-independent operator perturbatively or numerically, in a similar way as described in Eq. \eqref{blockdiagFloq} for periodic systems \cite{Shirley65}.
Specifically, it is argued \cite{MMFT} that finding the unitary transformation that relates $\tilde{\mathcal{K}}_{bd}$ and $\tilde{\cal{K}}_d$ is essentially equivalent to transforming $\mathcal{K}$ in Eq. \eqref{Kq} to  $\mathcal{K}_B$ in Eq. \eqref{KB}.
This method has then been applied in a variety of fields, leading to successful results \cite{Chu04,MMFTii,MMFTiii,MMFTiv,Son08,Forney10,Zhao15}.

In the next section we will give an explanation why, despite arguing that the proof of MMFT is not entirely rigorous and possibly incomplete,
this method can still lead to valid results.
We shall do this without imposing any intermediate periodicity in the system but rather by directly defining an extended Hilbert space, in an analogous way as described in Sec. \ref{sec:timeindep} for periodic systems.

\section{Quasi-periodic reducibility in Fourier space}\label{sec:QPFourier}

The possibility to express the time-evolution operator $U(t)$ of quasi-periodically driven systems in a generalized Floquet decomposition can be formulated in terms of the reducibility of the Schr\"odinger equation, as described in Sec. \ref{sec:timeindep} for periodic systems.  In the quasi-periodic case, 
we seek a quasi-periodic unitary $U_Q(t)$ that transforms the operator $K(t)=H(t)-i\partial_t$ to
\begin{eqnarray}\label{Ktransqp}
U_Q(t)K(t)U_Q^\dagger(t)=H_Q-i\partial_t,
\end{eqnarray}
where $H_Q=U_Q(t){H}(t)U_Q^\dagger(t)-iU_P(t)(\partial_tU_Q^\dagger(t))$ is a time-independent Hermitian operator and $U_Q(0)=\id$ 
\footnote{Similarly to the operator $H_F$ introduced in Eq. \eqref{Ktrans}, the eigenvalues of $H_Q$ are only defined up $\textbf{n}\cdot\boldsymbol{\omega}$, with a vector of integers $\textbf{n}$ and the frequency vector $\boldsymbol{\omega}$ of the Hamiltonian $H(t)$.}. 

In Sec. \ref{sec:timeindep}, we have described how the transformation in Eq. \eqref{Ktrans}---which is known to exist due to Floquet's theorem---can be solved within a time-independent formalism using Fourier series. Here, we expand this formalism to include quasi-periodic systems and  show how the transformation in Eq. \eqref{Ktransqp} can be similarly formulated in terms of the block-diagonalization of a time-independent operator.
With this, we will not aim at proving the existence of the decomposition of the time-evolution operator in Eq. \eqref{decompqp}, but rather assume its existence and construct the corresponding effective Hamiltonian.

Similarly as described in Sec. \ref{sec:Floq} for periodic systems, the Fourier coefficients of quasi-periodic states can be defined as the  Fourier components of vectors in $\mathcal{H}\otimes L^2(\mathbb{T}^d)$, where $\mathcal{H}$ is the original system's Hilbert space and $L^2(\mathbb{T}^d)$ is the space of square-integrable functions on a $d$-dimensional torus.
Thereafter, the isomorphism between the space $L^2(\mathbb{T}^d)$ and the sequence space $l^2(\mathbb{Z}^d)$ can be employed to work within a time-independent or Fourier formalism.
In this manner, quasi-periodic operators $B(t)=\sum_\textbf{n} B_\textbf{n}e^{i\textbf{n}\cdot\boldsymbol{\omega} t}$ can be mapped to
\begin{eqnarray}
\mathcal{B}=\sum_\textbf{n} B_\textbf{n}\otimes \sigma_\textbf{n},
\end{eqnarray}
were the ladder operators $\sigma_\textbf{n}=\sum_{\textbf{m}}|\textbf{m}+\textbf{n}\rangle\langle \textbf{m}|$ are defined in terms of a basis $|\textbf{n}\rangle$ of the sequence space $l^2(\mathbb{Z}^d)$ and satisfy $\sigma_\textbf{n}|\textbf{m}\rangle=|\textbf{n}+\textbf{m}\rangle$. Similarly, the derivative operator $-i\partial_t$ can be mapped to
\begin{eqnarray}
\mathcal{D}=\id\otimes \hat{\textbf{n}}\cdot\boldsymbol{\omega},
\end{eqnarray}
with the number operator $\hat{\textbf{n}}=\sum_\textbf{n} \textbf{n}|\textbf{n}\rangle\langle \textbf{n}|$ satisfying $\hat{\textbf{n}}|\textbf{n}\rangle=\textbf{n}|\textbf{n}\rangle$ and the commutation relation $[\hat{\textbf{n}},\sigma_\textbf{m}]=\textbf{m}\sigma_\textbf{m}$. The operator $K(t)=H(t)-i\partial_t$ can then be associated to the operator
\begin{eqnarray}
\mathcal{K}=\sum_\textbf{n} H_\textbf{n}\otimes \sigma_\textbf{n} +\id\otimes \hat{\textbf{n}}\cdot \boldsymbol{\omega},
\end{eqnarray}
which coincides with the operator already introduced in Eq. \eqref{Kq}.
In this way, the transformation in Eq. \eqref{Ktransqp} is then given by
\begin{eqnarray}\label{Qblockdiag}
\mathcal{K}_B=\mathcal{U}_Q\mathcal{K}\mathcal{U}_Q^\dagger=H_Q\otimes\id+\id\otimes \hat{\textbf{n}}\cdot \boldsymbol{\omega},
\end{eqnarray}
where the transformation $\mathcal{U}_Q$ has the form
\begin{eqnarray}\label{UqFloq}
\mathcal{U}_Q=\sum_\textbf{n} U_\textbf{n}\otimes \sigma_\textbf{n},
\end{eqnarray}
since it describes a quasi-periodic unitary.

The block-diagonalization described in Eq. \eqref{Qblockdiag} offers an alternative formulation of the transformation in Eq. \eqref{Ktransqp}, and indeed coincides with the transformation relating $\tilde{\mathcal{K}}_{bd}$ and $\tilde{\mathcal{K}}_d$ defined in Eqs. \eqref{Ktensor} and
\eqref{eq:Ktransform}.
That is, even if the general premise of MMFT is not satisfied, its explicit application is still correct as long as reducibility holds.

\section{Generalized Floquet-Magnus expansion}\label{sec:High}

General results of reducibility for first-order differential equations with quasi-periodic coefficients are not to be expected \cite{Eliasson1998,eliasson2006linear}, but the situation is better understood 
if the driving amplitude is small as compared to the norm $|\boldsymbol{\omega}|$ of the frequency vector. Using unitless variables $\tau=|\boldsymbol{\omega}|t$, which are common in the mathematical literature, the corresponding differential equation reads
\begin{equation}
i\partial_\tau U(\tau)=\frac{H(\tau)}{|\boldsymbol{\omega}|}U(\tau)\ .
\end{equation}
The regime of the new rescaled Hamiltonian, which is quasi-periodic with frequencies $\boldsymbol{\omega}/|\boldsymbol{\omega}|$, is referred to a {\em close-to-constant}, whereas the term {\em fast driving} is more common in the physics literature.
In this regime and under suitable hypothesis of regularity, non-degeneracy and strong nonresonance of the frequencies,  reducible and non-reducible systems are mixed like Diophantine and Liouvillean numbers: most systems are reducible but non-reducible ones are dense \cite{eliasson:so3, krikorian:annals,Fraczek2004OnSu2,karaliolios:global}. 
Morover,
the generalized Floquet decomposition of the time-evolution operator in Eq. \eqref{decompqp} can be found with any given accuracy, for $|\boldsymbol{\omega}|^{-1}$ sufficiently small, provided it exists \cite{Simo94,Treshchev1995,Jorba1997}.
In practice, this is often done through an expansion in terms of powers of $|\boldsymbol{\omega}|^{-1}$.

In this section, we will derive a generalization of the Floquet-Magnus expansion \cite{Casas01,Blanes09,Magnus54}  to quasi-periodic systems and provide a perturbative exponential expansion of the time-evolution operator with the desired Floquet representation. 
This will allow us to identify $H_Q$ as the effective Hamiltonian that captures the dynamics of the system in a suitable fast-driving regime.

We start the derivation by reproducing the steps of the regular Floquet-Magnus expansion \cite{Casas01} and introducing the desired decomposition of the time-evolution operator
\begin{eqnarray}\label{qpdecomp2}
U(t)=U_Q^\dagger(t)e^{-iH_Qt}
\end{eqnarray} 
into the Schr\"odinger equation $i\partial_t U(t)=H(t)U(t)$, which yields the differential equation
\begin{eqnarray}\label{diffeq1}
i\partial_t U_Q^\dagger(t)=H(t)U_Q^\dagger(t)-U_Q^\dagger(t)H_Q.
\end{eqnarray}
Then, we define the quasi-periodic Hermitian operator $Q(t)$ as the generator of the quasi-periodic unitary $U_Q(t)$ via the relation
\begin{eqnarray}\label{QPGen}
U_Q(t)=e^{iQ(t)}.
\end{eqnarray}
Introducing the expression in Eq. \eqref{QPGen} into Eq. \eqref{diffeq1} and using a power series expansion for the differential of the exponential \cite{Magnus54,Wilcox67,Casas01}, one obtains the non-linear differential equation \cite{Casas01}
\begin{eqnarray}\label{adeq}
\partial_t Q(t)=\sum_{k=0}^\infty \dfrac{B_k}{k!}(-i)^k \mbox{ad}^k_{Q(t)}\left(H(t)+(-1)^{k+1}H_Q\right),
\end{eqnarray}
where $B_k$ denote the Bernoulli numbers and $\mbox{ad}$ is the adjoint action defined via $\mbox{ad}^k_A B=[A,\mbox{ad}^{k-1}_A B]$ for $k\geq1$ and $\mbox{ad}^0_A B=B$.

The next step in the derivation is to consider a series expansion for the operators $H_Q$ and $Q(t)$ of the form
\begin{eqnarray}\label{HFexpansion}
H_Q&=&\sum_{n=1}^\infty H_Q^{(n)}\\
Q(t)&=&\sum_{n=1}^\infty Q^{(n)}(t), \label{Gexpansion}
\end{eqnarray}
with $Q^{(n)}(0)=0$ and where the superscript indicates the order of the expansion. 
After introducing the series in Eqs. \eqref{HFexpansion} and \eqref{Gexpansion} into Eq. \eqref{adeq}  and equating the terms with the same order, one obtains the differential equation
\begin{eqnarray}\label{Diffeq2}
\partial_t Q^{(n)}(t)&=&A^{(n)}(t)-H_Q^{(n)},
\end{eqnarray}
with $A^{(1)}(t)=H(t)$ and 
\begin{eqnarray}\label{A}
A^{(n)}(t)&=&\sum_{k=1}^{n-1}\dfrac{B_k}{k!}(X_k^{(n)}(t)+(-1)^{k+1}Y_k^{(n)})
\end{eqnarray}
for $n\geq2$. 
The operators $X_k^{(n)}(t)$ and $Y_k^{(n)}(t)$ in Eq. \eqref{A} are given recursively by
\begin{eqnarray}
X_k^{(n)}(t)&=&\sum_{m=1}^{n-k}[Q^{(m)}(t),X_{k-1}^{(n-m)}(t)]\\
Y_k^{(n)}(t)&=&\sum_{m=1}^{n-k}[Q^{(m)}(t),Y_{k-1}^{(n-m)}(t)]
\end{eqnarray}
for $1\leq k\leq n-1$, with $X_0^{(1)}=-iH(t)$, $X_0^{(n)}=0$ for $n\geq2$, and $Y_0^{(n)}=-iH_Q^{(n)}$ for all $n$. 

An important feature of the differential equation in Eq. \eqref{Diffeq2} is the structure of the operator $A^{(n)}(t)$, which only contains terms involving the Hamiltonian $H(t)$ or operators $Q^{(m)}(t)$ and $H_Q^{(m)}$ of a lower order, i.e. with $m<n$. This allows to solve Eq. \eqref{Diffeq2}  by just integrating the right hand side of the equation, which leads to
\begin{eqnarray}\label{Qn}
Q^{(n)}(t)&=&\int_0^t \left(A^{(n)}(t)-H_Q^{(n)}\right)dt.
\end{eqnarray}
Moreover, even though Eq. \eqref{Diffeq2} describes a differential equation for $Q^{(n)}(t)$, the solutions for both $Q^{(n)}(t)$ and $H_Q^{(n)}$ can be inferred from it by imposing suitable conditions on the time dependence of $Q^{(n)}(t)$. In the periodic case, for example, the operators $H_Q^{(n)}$ are fixed by the requirement that $Q^{(n)}(t)$ is a periodic operator \cite{Casas01}. 

In the quasi-periodic case, we can determine $H_Q^{(n)}$ by exploiting the quasi-periodicity of  $Q^{(n)}(t)$ and  $A^{(n)}(t)$.
This essentially results from the fact that, in order for the integral of a quasi-periodic operator $O(t)=\sum_\textbf{n} O_\textbf{n} e^{i\textbf{n}\cdot\boldsymbol{\omega}t}$ to be quasi-periodic, it must satisfy that $O_\textbf{0}=\lim_{T\to \infty}\frac{1}{T}\int_0^T O(t) dt=0$.
As a consequence, in order for $Q^{(n)}(t)$ in Eq. \eqref{Qn} to be quasi-periodic, $H_Q^{(n)}$ must read
\begin{eqnarray}\label{HQn}
H_Q^{(n)}&=&\lim_{T\to \infty}\dfrac{1}{T}\int_0^T A^{(n)}(t) dt.
\end{eqnarray}
Eqs. \eqref{Qn} and \eqref{HQn} can be solved for any $n>1$ provided that the solutions for $m<n$ are known. Since they can be readily solved for $n=1$, Eqs. \eqref{Qn} and \eqref{HQn} thus contain the necessary information to recursively derive all the terms in the expansions of $Q(t)$ and $H_Q$ in Eqs. \eqref{HFexpansion} and \eqref{Gexpansion}, respectively.

After performing the integrations in Eqs. \eqref{Qn} and \eqref{HQn}, the first two terms terms of the series for $H_Q$ become
\begin{eqnarray}\label{HQ1}
H_Q^{(1)}&=&H_\textbf{0}\\
H_Q^{(2)}&=&\dfrac{1}{2}\sum_{\textbf{n}\neq 0}\dfrac{[H_\textbf{n},H_{-\textbf{n}}]}{\boldsymbol{\omega}\cdot \textbf{n}}+\sum_{\textbf{n}\neq 0}\dfrac{[H_\textbf{0},H_{\textbf{n}}]}{\boldsymbol{\omega}\cdot \textbf{n}},\label{HQ2}
\end{eqnarray}
where $H_\textbf{n}$ are the Fourier coefficients of the quasi-periodic Hamiltonian, as defined in Eq. \eqref{qpham}. Similarly, the first two terms of $Q(t)$ read
\begin{eqnarray}\label{Q1}
Q^{(1)}(t)&=&-i\sum_{\textbf{n}\neq \textbf{0}} \dfrac{H_\textbf{n}}{\textbf{n}\cdot \boldsymbol{\omega}}(e^{i\textbf{n}\cdot \boldsymbol{\omega} t }-1) \\
Q^{(2)}(t)&=&\dfrac{i}{2}\sum_{\textbf{n}\neq \textbf{0}} \dfrac{[H_\textbf{0},H_\textbf{n}]}{(\textbf{n}\cdot \boldsymbol{\omega})^2}(e^{i\textbf{n}\cdot \boldsymbol{\omega} t }-1)\nonumber \\
&&+\dfrac{i}{2}\sum_{\textbf{n}\neq \textbf{0} ; \textbf{m}\neq -\textbf{n}}\dfrac{[H_\textbf{n},H_\textbf{m}]}{\textbf{n}\cdot \boldsymbol{\omega} \  (\textbf{n}+\textbf{m})\cdot \boldsymbol{\omega}}(e^{i(\textbf{n}+\textbf{m})\cdot \boldsymbol{\omega} t }-1)\nonumber \\
&&+\dfrac{i}{2}\sum_{\textbf{n}\neq \textbf{0} ; \textbf{m}\neq \textbf{0}}\dfrac{[H_\textbf{n},H_\textbf{m}]}{\textbf{n}\cdot \boldsymbol{\omega} \  \textbf{m}\cdot \boldsymbol{\omega}}(e^{i\textbf{m}\cdot \boldsymbol{\omega} t }-1)\ .
\label{Q2}
\end{eqnarray}  
Consistently with the periodic case, the results obtained here reduce to the regular Floquet-Magnus expansion when the frequency vector $\boldsymbol{\omega}$ contains only one element. Moreover, by using the Baker-Campbell-Hausdorff formula \cite{Varadarajan}, one can verify that the expressions in Eqs. \eqref{HQ1}-\eqref{Q2} coincide with the first terms of the regular Magnus expansion \cite{Magnus54}, which applies for general time-dependent systems.
This formal expansion can also be linked \cite{Chartier2012AEstimates} to the method of averaging for quasi-periodic systems \cite{Simo94} to obtain exponentially small error estimates in the quasi-periodic case.

The expansion of the operators $H_Q$ and $Q(t)$ introduced in Eqs. \eqref{HFexpansion} and \eqref{Gexpansion} can be interpreted as a series expansion in powers of $|\boldsymbol{\omega}|^{-1}$ such that, in a suitable  fast-driving regime,  the lowest-order terms of the series are the most relevant to describe the dynamics of the system.  Even though the convergence of the quasi-periodic Floquet-Magnus expansion is in general not guaranteed and requires further investigations, this permits us to identify effective Hamiltonian analogously as done for periodic systmes.

In fast-driving regimes where the fundamental driving frequencies are the largest energy scales of the system, the two unitaries $U_Q(t)$ and $e^{-iH_Qt}$ of the time-evolution operator in Eq. \eqref{qpdecomp2}  capture two disctinct behaviours of the system's dynamics. On the one hand, the unitary $U_Q(t)$ describes fast quasi-periodic fluctuations dictated by the fast frequencies $\boldsymbol{\omega}$. On the other hand, the operator $e^{-iH_Qt}$ captures the slower dynamics of the system
characterized by the internal energy scales of  $H_Q$, which can be thus identified as the effective Hamiltonian of the system.

\section{Quasi-periodically driven Lambda system}
\label{sec:Lambda}

\begin{figure*}[t] 
  \centering
\small{(a)} \includegraphics[width=0.43\textwidth]{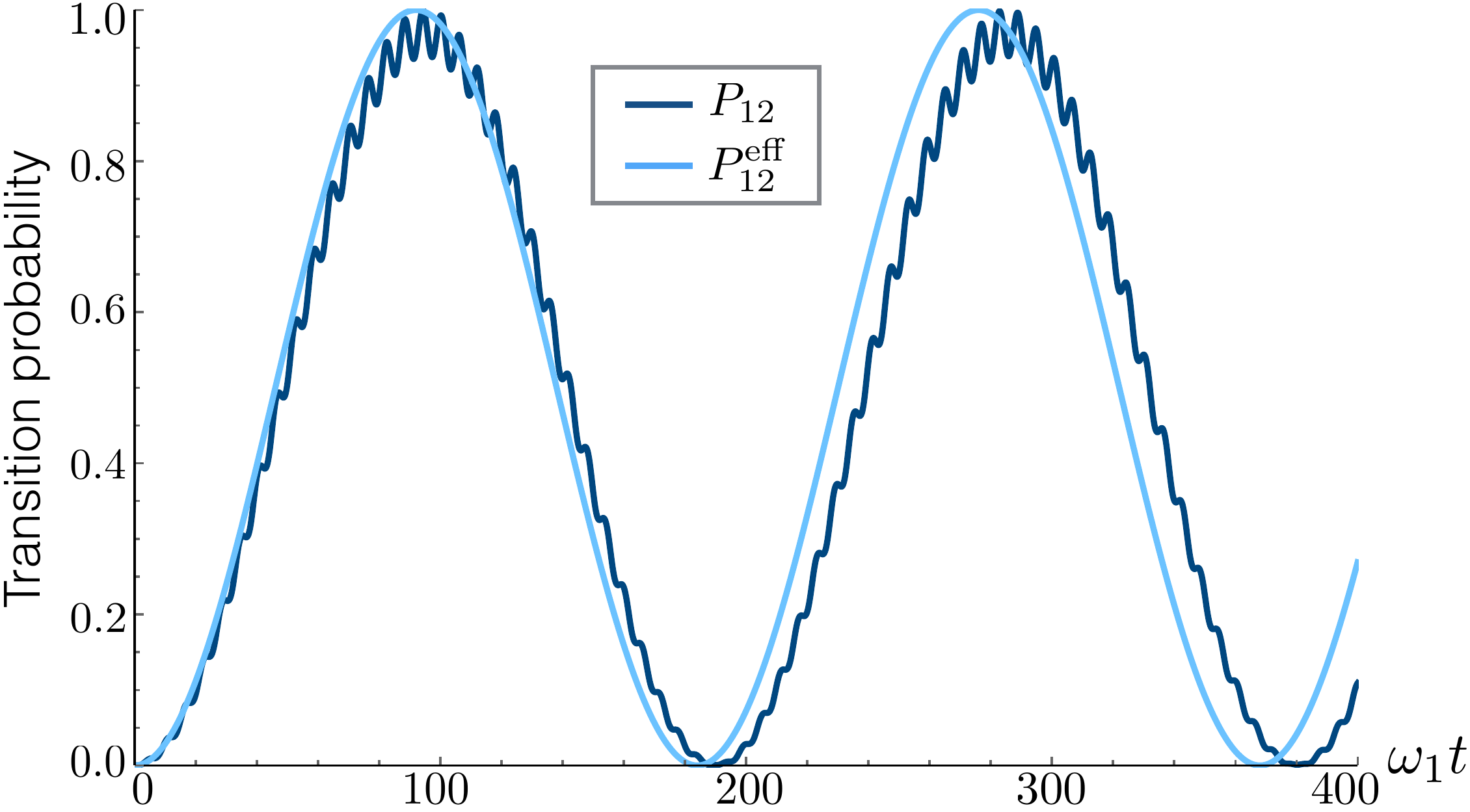}  \hspace{1cm}   \small{(b)} \includegraphics[width=0.43\textwidth]{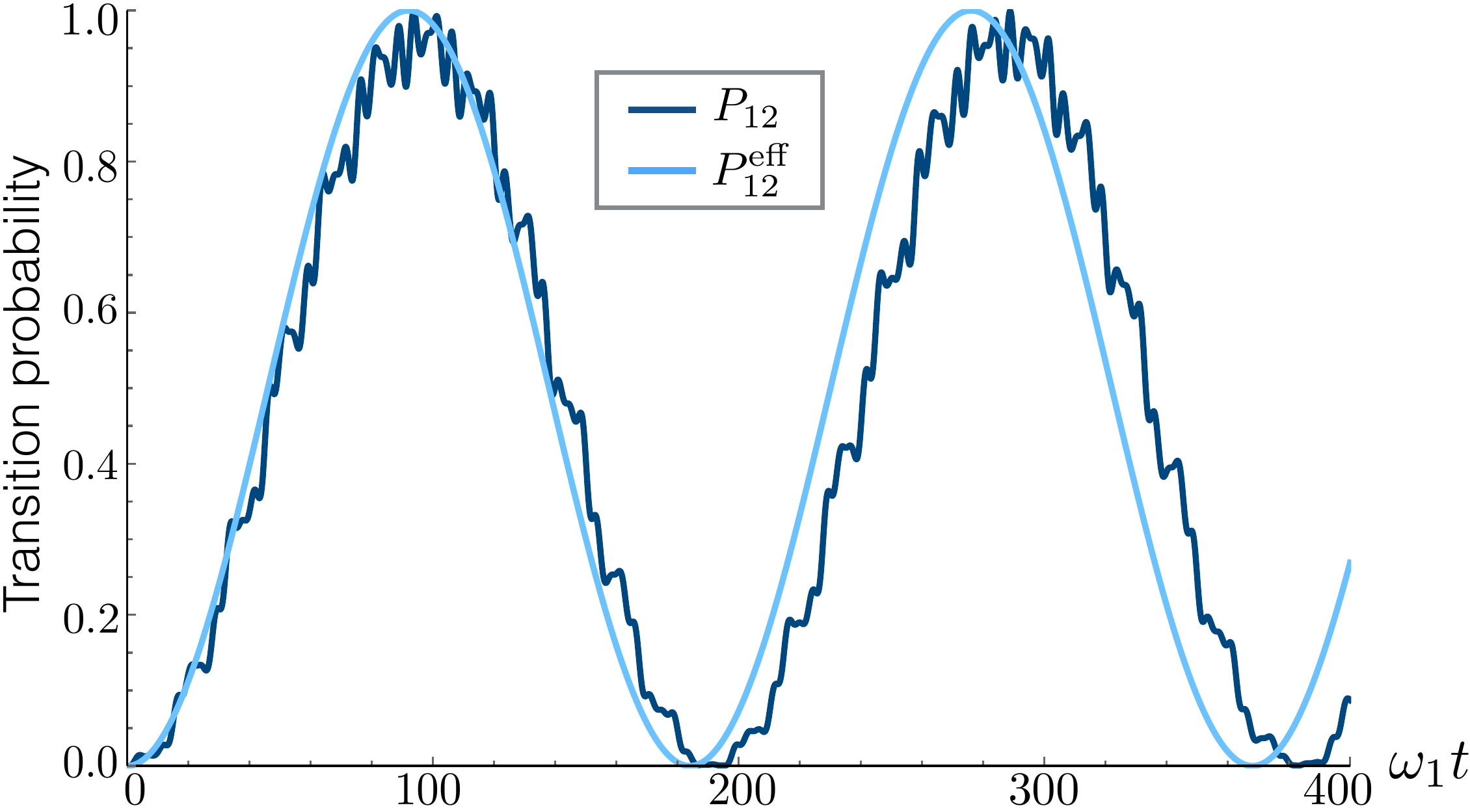}
  \caption{Comparison between the exact and effective transition probabilities $P_{12}(t)=|\langle 1|U(t)|2\rangle|^2$ and $P_{12}^{\rm eff}(t)=|\langle 1|U_{\rm eff}(t)|2\rangle|^2$ for the periodically (a) and quasi-periodically (b) driven Lambda system. In (a), a periodic driving function  $f(t)=\Omega e^{i\omega_1 t}$  with $\Omega/\omega_1=0.1(1+\sqrt{2}/2)^{1/2}$ is considered. In (b), the results correspond to a quasi-periodic function  $f(t)=\Omega( e^{i\omega_1 t}+ e^{i\sqrt{2}\omega_1 t})$ with $\Omega/\omega_1=0.1$. 
  The parameters of the driving function in (a) and (b) are such that they lead to the same effective rate $\Omega_{\rm eff}$ in Eq. \eqref{Omegaeff}.
\label{Fig:comp}}
\end{figure*}

In this section, we will illustrate with a quasi-periodically driven Lambda system the possibility to approximate the dynamics of quasi-periodically driven systems in terms of a truncation of the effective Hamiltonian $H_Q$.

The Lambda system describes an atomic three energy-level system with two  ground states $|1\rangle$ and $|2\rangle$ that are coupled to an excited state $|3\rangle$ via a time-dependent laser field. The time-dependent coupling allows one to indirectly mediate a transition between the states $|1\rangle$ and $|2\rangle$ without significantly populating the excited state and, in this way, overcome the impossibility to drive a direct transition between the two degenerate ground states.
This method also permits the implementation of non-trivial phases in the tunneling rate of particles \cite{Cheuk12,Wang12} and constitutes a building block in many quantum simulations \cite{Lin09,Lin11,Aidelsburger13,Aidelsburger15}.

The Hamiltonian of the Lambda system in an interaction picture reads
\begin{eqnarray}\label{Hlambda}
H(t)= f(t)|3\rangle \left( \langle 1|+\langle 2|\right)+\mbox{H.c},
\end{eqnarray}
where $f(t)$ is usually a periodic function but, here,  we consider it to be quasi-periodic; i.e. of the form
\begin{eqnarray}
f(t)=\sum_\textbf{n}f_\textbf{n}e^{i\textbf{n}\cdot\boldsymbol{\omega}t},
\end{eqnarray}
with a frequency vector $\boldsymbol{\omega}$ and Fourier components $f_\textbf{n}$.
Moreover, we require the static Fourier component to vanish, i.e. $f_\textbf{0}=0$, in order to ensure that the dominant dynamics of the system do not yield transitions between the ground states and the excited state.

With the Hamiltonian of the quasi-periodically driven Lambda system in Eq. \eqref{Hlambda}, the first two terms of the effective Hamiltonian expansion in Eqs. \eqref{HQ1} and \eqref{HQ2}  become $H_Q^{(1)}=0$ and
\begin{eqnarray}
H_Q^{(2)}=\Omega_{\rm eff}\left( (|1\rangle+|2\rangle)(\langle1|+\langle2|)-2|3\rangle\langle3|\right),
\end{eqnarray}
respectively. As the first term vanishes, the leading order term of the effective Hamiltonian is thus  given by $H_Q^{(2)}$, which describes  transitions between the ground states of the system with a rate
\begin{eqnarray}\label{Omegaeff}
\Omega_{\rm eff}=\sum_\textbf{n} \frac{|f_\textbf{n}|^2}{\textbf{n}\cdot\boldsymbol{\omega}}.
\end{eqnarray}

\begin{figure}[t]
\includegraphics[width=0.45\textwidth]{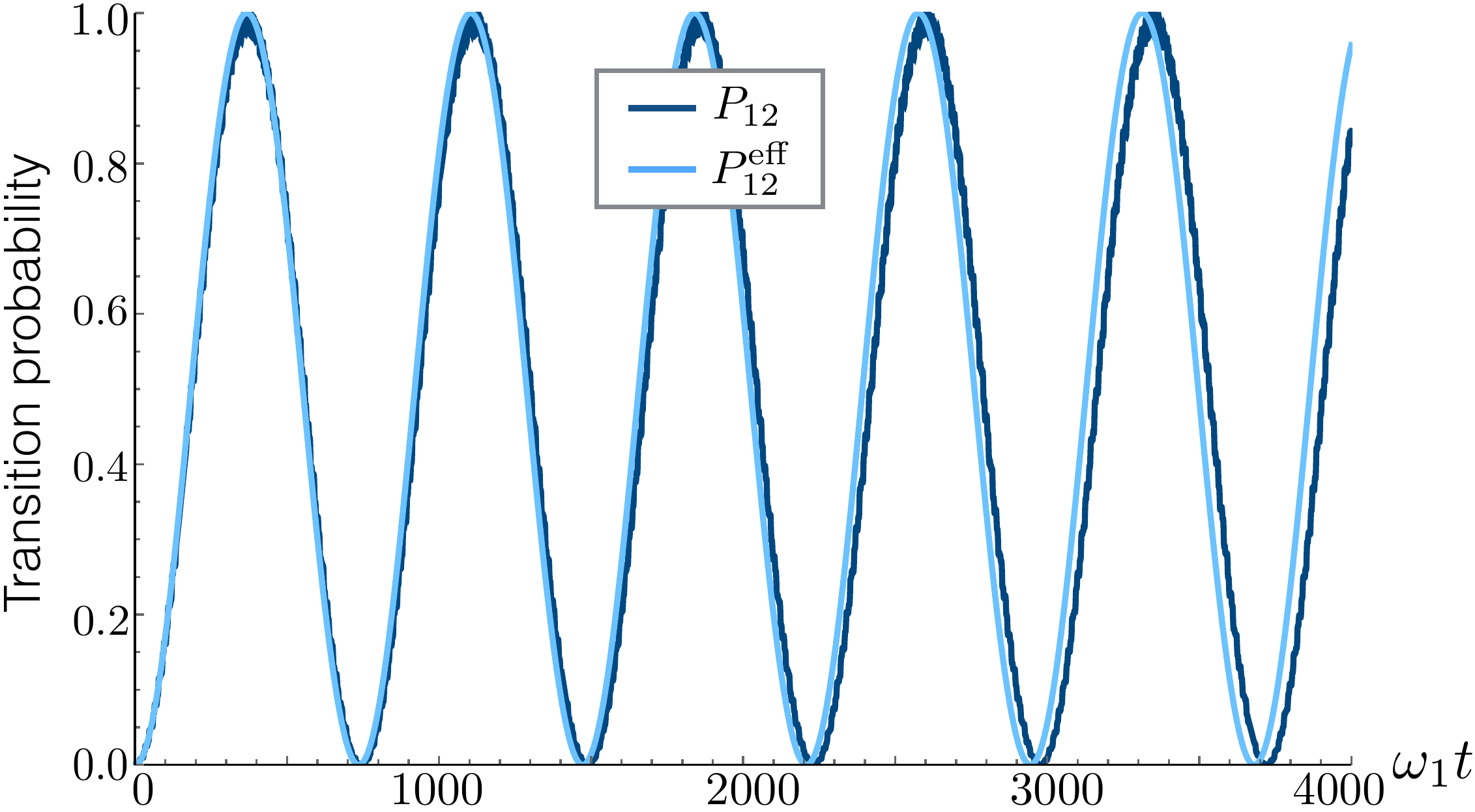}
  \caption{Plot of the transition probabilities $P_{12}(t)=|\langle 1|U(t)|2\rangle|^2$ and $P_{12}^{\rm eff}(t)=|\langle 1|U_{\rm eff}(t)|2\rangle|^2$ as a function of time for a quasi-periodic Lambda system with $f(t)=\Omega( e^{i\omega_1 t}+ e^{i\sqrt{2}\omega_1 t})$ and $\Omega/\omega_1=0.05$. \label{Fig:Lambda}}
\end{figure}

In order to illustrate the possibility to approximate the dynamics of the system in terms of a truncation of $H_Q$, we compare in Figs. \ref{Fig:comp} and \ref{Fig:Lambda} matrix elements of a numerically exact calculation of the time-evolution operator of the system $U(t)$ and the effective time-evolution operator
\begin{equation}\label{Ueff}
U_{\rm eff}(t)=e^{-iH_{\rm Q}^{(2)}t}
\end{equation}
for different driving functions $f(t)$. Specifically, we display the transition probabilities $P_{12}(t)=|\langle 1|U(t)|2\rangle|^2$ and $P_{12}^{\rm eff}(t)=|\langle 1|U_{\rm eff}(t)|2\rangle|^2$, which describe the exact and effective transitions between the ground states of the Lambda system.

In Fig. \ref{Fig:comp}, we compare the performance of the driven Lambda system for a periodic and quasi-periodic driving functions in a moderately fast driving regime. In Fig. \ref{Fig:comp} (a), a periodic driving is considered, which yields exact dynamics that exhibit fast regular fluctuations around the slower effective dynamics. On the contrary, we show in Fig. \ref{Fig:comp} (b) how a quasi-periodic driving leads to a pattern with seemingly erratic fluctuation around the effective dynamics.  In the regime where the fluctuations can be neglected, however, their regularity is irrelevant. This supports the view that, since quasi-periodic functions provide a more general parametrization of  the driving protocol, quasi-periodically driven quantum systems have the potential to expand the accessible effective dynamics in a variety of experimental setups \cite{Yuce13}.

Another aspect that is apparent from Fig. \ref{Fig:comp} is the drift between the exact and effective dynamics.
This is not a characteristic feature of quasi-periodically driven systems but rather results from the truncation of the operator $H_Q$. Including higher order terms in the expansion of  $H_Q$ or considering a faster driving regime, the approximation would be improved and the exact and effective dynamics of the system would better overlap for longer times. 
Indeed, in Fig. \ref{Fig:Lambda} we consider a quasi-periodic function with higher frequencies and observe that  the effective time-evolution operator in Eq. \eqref{Ueff} approximates better the exact dynamics of the system for longer times. This highlights the possibility to use the generalized Floquet-Magnus expansion derived in Sec. \ref{sec:High} in order to derive time-independent effective Hamiltonians that capture well the dynamics of quasi-periodic systems in a suitable fast-driving regime.

\section{Conclusions}\label{sec:Conc}

Despite concerted efforts towards generalisation of Floquet's theorem for quasi-periodic systems, there are still many open questions regarding the existence of Floquet-like decompositions.
Although a rigorous footing is not complete,  effective Hamiltonians can be constructed for weakly and/or rapidly driven quantum systems.

Since the restriction to periodic driving naturally imposes restrictions on the effective Hamiltonians that can be achieved, the use of quasi-periodic Hamiltonians is a promising route for quantum simulations.
The increased freedom in accessible time-dependencies makes quasi-periodic driving a highly interesting basis for the identification of accurate implementations of effective dynamics by means of optimal control.
As such, one can expect that quasi-periodic driving will find increased application in quantum simulations, and that the increased interest in quantum physics will trigger activities in mathematics towards the existence of Floquet-like decompositions and the convergence of perturbative expansions.

\begin{acknowledgments}
A.V. and F.M. acknowledge financial support by the European Research Council within the project ODYCQUENT. The research of J.P. is supported by grants 2014-SGR-00504 and MTM2015-65715-P.
\end{acknowledgments}

\appendix*

\section{The propagator in Floquet theory}
\label{appendix}

Here, we show that
\begin{eqnarray}\label{eq:app1}
\tilde{U}(t)=\sum_n \langle n |e^{-i\tilde{\mathcal{K}}t}|0\rangle e^{in\omega t},
\end{eqnarray}
as given in Eq.\eqref{UFloq} is indeed the propagator induced by $H(t)$,
{\it i.e.} that it satisfies the Schr\"odinger equation with the initial condition $\tilde{U}(0)=\id$.

The time-derivative of Eq.\eqref{eq:app1} reads
\begin{eqnarray}\label{eq:app2}
i\partial_t\tilde{U}(t)=\sum_n \langle n |
(\tilde{\mathcal{K}}-n\omega)e^{-i\tilde{\mathcal{K}}t}|0\rangle e^{in\omega t}\ .
\end{eqnarray}
Using the explicit form
\begin{eqnarray}
\tilde{\mathcal{K}}=\sum_n H_n\otimes \sigma_n +\id\otimes \omega \hat{n}.
\end{eqnarray}
and
\begin{eqnarray}
\langle n |\hat n=\langle n |n\ ,
\end{eqnarray}
Eq.~\eqref{eq:app2} reduces to
\begin{eqnarray}\label{eq:app3}
i\partial_t\tilde{U}(t)&=&\sum_{nm} \langle n | H_m\otimes \sigma_m e^{-i\tilde{\mathcal{K}}t}|0\rangle e^{in\omega t}\\
&=&\sum_{nm} H_m\langle n | \sigma_m e^{-i\tilde{\mathcal{K}}t}|0\rangle e^{in\omega t}\\
&=&\sum_{nm} H_m\langle n-m | e^{-i\tilde{\mathcal{K}}t}|0\rangle e^{in\omega t}\ .
\end{eqnarray}
Replacing the summation index $n$ by $n+m$ leads to
\begin{eqnarray}\label{eq:app4}
i\partial_t\tilde{U}(t)
&=&\sum_{nm} H_m\langle n | e^{-i\tilde{\mathcal{K}}t}|0\rangle e^{i(n+m)\omega t}\\
&=&\sum_{m} H_me^{im\omega t}\ \sum_n\langle n | e^{-i\tilde{\mathcal{K}}t}|0\rangle e^{in\omega t}\\
&=&
H(t)\tilde{U}(t)\ .
\end{eqnarray}
The initial condition $\tilde{U}(0)=\id$ results directly from $\langle n|e^{-i\tilde{\mathcal{K}}0}|0\rangle=\id$.

\bibliography{bibqp}

\end{document}